\documentclass[12pt,preprint]{aastex}

\lefthead{Bekki \&  Tsujimoto}
\righthead{Impact of high velocity clouds}

\begin{document}

\title{Origin of   the unusually low nitrogen abundances in young
populations of the Large Magellanic Cloud}

\author{Kenji Bekki} 
\affil{
ICRAR,
M468,
The University of Western Australia
35 Stirling Highway, Crawley
Western Australia, 6009 
}

\and

\author{Takuji Tsujimoto} 
\affil{
National Astronomical Observatory, Mitaka-shi, Tokyo 181-8588, Japan}

\begin{abstract}

It is a longstanding problem that HII regions and very young
stellar populations in the Large Magellanic Cloud (LMC)
have the nitrogen abundances
 ([N/H]) by a factor of $\sim 7$ lower than the solar value.
We here discuss  a new scenario in which
the observed unusually low nitrogen abundances
can be closely associated with recent collision and subsequent
accretion  of 
H~{\sc i}   high velocity clouds (HVCs) that surround the Galaxy and
have   low nitrogen abundances.
We show that if the observed low [N/H] is limited to 
very young stars with ages less than $\sim 10^7$ yr, 
then  the collision/accretion rate of the HVCs onto the LMC
needs to be $\sim 0.2 {\rm M}_{\odot}$ yr$^{-1}$ 
(corresponding to the total HVC mass of $10^6-10^7 {\rm M}_{\odot}$)
to dilute
the original interstellar medium (ISM) before star formation.
The required accretion rate means that even if the typical mass
of HVCs accreted onto the LMC is $\sim 10^7 {\rm M}_{\odot}$,
the Galaxy needs to have  $\sim 2500$ massive  HVCs within the LMC's
orbital radius with respect to the Galactic center. 
The required rather large number of massive HVCs drives us to 
suggest that 
the HVCs are not likely  to efficiently dilute the ISM
of the LMC and consequently lower the  [N/H]. 
We thus suggest 
the transfer of gas with low [N/H] from the Small Magellanic Cloud 
(SMC) to the LMC as  a promising  scenario that can explain
the observed low [N/H].

\end{abstract}

\keywords{
Galaxy: halo --
(galaxies): Magellanic Clouds --
galaxies: evolution -- 
galaxies: halos -- 
galaxies: abundances
}

\section{Introduction}

Chemical abundances in gaseous and stellar components
of the LMC and the SMC have provide
many clues to formation and evolution of the Clouds
(e.g.,  Da Costa \& Hatzidimitriou 1998;
Pagel \&  Tautvaisiene 1998; Venn 1999;
Piatti et al. 2001; Hill 2004; 
Cole et al. 2005;
Carrera et al. 2008; Tsujimoto \& Bekki 2009). 
One of intriguing  observational results  in terms of 
chemical abundances of the LMC is
that star-forming HII regions and very young stellar populations
in the LMC show apparently low [N/H] that is by a factor
of $6-7$ lower than the solar value 
(e.g., Korn et al. 2002; Hill 2004; van Loon et al. 2010)
while other elements such as O and Ne are underabundant by a factor of $\sim 2$
(Hill 2004).
It remains observationally  unclear whether only these young components
in the LMC have such low [N/H]  or  other older ones also have it.
It would be theoretically  unlikely that all stellar populations and gas
in the LMC have such low [N/H].

         In any standard chemical evolution models,
the N abundance will increase with time and reach the same level of
enrichment as those of other elements in the reasonable scheme that 
stellar ejecta from asymptotic giant branch (AGB) stars
(which are considered to be the major
production site for N) in a galaxy is well recycled into
and mixed with the ISM and subsequently the mixed gas
is efficiently  converted into new stars.
Therefore the observed  low [N/H]  $\sim -0.8$
in HII regions and very young
stellar populations  of  {\it the present} LMC
with [Fe/H] $\sim-0.3$
appears to be at odds with theoretical predictions
of previous chemical evolution models 
(e.g., Henry et al. 2000;
Moll\'a et al. 2006).
One of possible ways to solve this
is to dilute the nitrogen abundance through recent infall of gas with
low [N/H] onto the LMC based on the assumption that
only young components
in the LMC have such low [N/H].
There would be a number of possibilities for the gas resource,
such as the Galactic HVCs,  the SMC, and some gas-rich dwarfs
orbiting the Galaxy.

The purpose of this paper is to discuss 
a new scenario in which
the observed low [N/H] can
be closely associated with the recent infall of
the Galactic HVCs onto the LMC.
Given that some HVCs presumably within the Galactic halo
are observed to have low [N/H] 
ranging from $-2.0$ to $-1.2$ (Collins et al. 2007),
the proposed scenario appears to be reasonable. 
In addition, the similarity of N/O ratio between 
the H II region ($\log{\rm (N/O)} \sim -1.5$) and
the HVCs ($\log{\rm (N/O)} \sim -1.6$) 
implies some connection between them.
Using an idealized model, 
we investigate how much accretion of the  HVCs is necessary
to dilute the star-forming ISM to the extent that [N/H]
of new stars can have the observed low [N/H].
We then estimate the possible number and total mass of the Galactic HVCs
for the required accretion rate of the HVCs.
Given that previous numerical simulations demonstrated (i) the accretion of HVCs onto
the Galaxy (e.g., Cameron \& Torra 1994) and (ii) possible
presence of a large number of HVCs in the Galactic halo
(e.g., Peek et al. 2008),   
the HVCs can be accreted onto the LMC if they collide with the LMC.

In the present paper, we do not intend to discuss why and how
the HVCs have very low [N/H] (e.g., Collins et al. 2007),
because this is simply beyond the scope of this paper. 
The origin of the low [N/H] would be closely associated with
tidal stripping of outer HI gas of ancient gas-rich dwarfs,
where chemical evolution did not proceed effectively 
(Bekki \& Tsujimoto 2010, in preparation): our future papers will
discuss why the stripped HI gas can have low [N/H] in detail.
This paper thus focuses on (i) whether gas infall from the HVCs
can explain the origin of the observed low [N/H] in the LMC
and (ii)  what other scenarios can explain it
if the HVC infall scenario is not so  plausible.

The Appendix in this paper shows the time evolution of [N/O]
of stars in the LMC based on the standard one-zone chemical evolution
model for the LMC and compares the simulated [N/O] of the present
LMC with the observed one. 
The Appendix thus can help readers to understand
(i) that this is a serious inconsistency between the theoretical prediction of [N/O]
for the present stellar populations of the LMC
and the corresponding observed one
and thus (ii) that other factors need to be considered to reproduce
the observed low nitrogen abundance of stars in the present LMC.

\section{The required accretion rate of the Galactic  HVCs}

We here assume that (i) all of young stellar populations with ages
less than $t_{\rm sf}$ have unusually low [N/H] in the LMC and (ii)
the populations are formed exclusively from mixed gas of the original
ISM of the LMC and the accreted gas (i.e., the Galactic HVCs).
We adopt the above first assumption, because previous observations show no clear
evidence for the presence of young stars and HII regions with normal [N/H]
(e.g., Russell \& Dopita 1990).  The above second assumption means that
the original ISM of the LMC has a ``normal'' [N/H] before external accretion
of gas and thus needs to be diluted by the accreted gas to have low [N/H].
We discuss how the present results change if we relax these model
assumption in \S 4.

We try to derive the total mass of the accreted HVCs ($M_{\rm HVC}$),
for $t_{\rm sf}$,
the observed  present star formation rate of the LMC ($R_{\rm sf}$),
the nitrogen abundance of the original ISM of the LMC ($A_{\rm g}$),
that of the HVCs ($A_{\rm HVC}$), 
that of young  population observed in the LMC ($A_{\rm obs}$),
star formation efficiency for the mixed gas (${\epsilon}_{\rm sf}$),
and the total mass of the original ISM that can be converted into
new stars ($M_{\rm g}$: therefore this is not the total mass
of the entire ISM of the LMC, it is the local gas mass (initially 
in  the LMC)
mixing with
the accreted HVCs to form new stars with low [N/H]).
We assume that $t_{\rm sf}$ is constant during 
the formation of young stars (i.e., for $\sim 10^7$ yr in most
models), because there is no/little observational evidence which
supports rapid change in star formation within an order of $10^7$ yr:
the periodic bursts of star formation are observationally suggested
(e.g., Harris et al. 2009), but they are  estimated for a time span
of 100 Myr to several Gyrs.

We use the following  two sets of equation to derive $M_{\rm HVC}$:
\begin{equation}
{\epsilon}_{\rm sf}(M_{\rm HVC}+M_{\rm g})
=t_{\rm sf}R_{\rm sf}
\end{equation}

and 

\begin{equation}
\frac{ A_{\rm HVC}M_{\rm HVC}+A_{\rm g}M_{\rm g} }
{ M_{\rm HVC}+M_{\rm g} }=A_{\rm obs}.
\end{equation}

For the above equation (1), we consider that
(i) the star formation rate estimated from HII regions of the LMC
($\sim 0.26$ ${\rm M}_{\odot}$ yr$^{-1}$; Kennicutt et al. 1995) is reasonable in the present study
and (ii) young stellar populations with ages
less than $t_{\rm sf}$ were continuously  formed with the star formation
rate of $R_{\rm sf}$.
We choose $A_{\rm obs}$ corresponding to the observed [N/H]
($=-0.8$) for HII regions and B-type of stars in the LMC
(Korn et al. 2002).
It is observationally unclear what the typical
value of $A_{\rm HVC}$ is,
though some observations show very low [N/H] 
ranging from $-2.0$ to $-1.2$ (Collins et al. 2007).
Therefore we consider that $A_{\rm HVC}$ is a  parameter
with the above observed range.
Since the present [Fe/H] of the LMC is observed to be $-0.3$
(e.g., van den Bergh 2000), we set the gaseous [N/H] to be $-0.3$
assuming [N/Fe]=0:
we use $A_{\rm g}$ corresponding to this [N/H] value.
We consider that $t_{\rm sf}\sim10^7$ yr is reasonable, because
the ages of stars (e.g., main-sequence B-type stars) observed
for estimation of N abundances 
(Korn et al. 2002) correspond roughly to the above
$t_{\rm sf}$. We however investigate models with 
different  $t_{\rm sf}$

Fig. 1 shows how $M_{\rm HVC}$ required to decrease 
[N/H] of the ISM to $\sim-0.8$  depends on the nitrogen abundance
of HVCs (denoted as ${\rm  [N/H]}_{\rm HVC}$) for $t_{\rm sf}=10^7$yr
in three models with different ${\epsilon}_{\rm sf}$.
The required $M_{\rm HVC}$ ranges from 
$\sim 1.8 \times 10^6  {\rm M}_{\odot}$  
to  $\sim2.2 \times 10^7 {\rm M}_{\odot}$
and is larger for smaller ${\epsilon}_{\rm sf}$ for
a given ${\rm  [N/H]}_{\rm HVC}$ (i.e., $A_{\rm HVC}$). 
The reason for this is as follows: Only a smaller fraction
of the accreted HVCs can be converted into new stars after mixing
with the original ISM in the models with a smaller ${\epsilon}_{\rm sf}$.
 Therefore 
a larger amount of HVCs needs to be accreted to form the observed
total mass of young stars with low [N/H].
The required $M_{\rm HVC}$  is larger for larger
${\rm  [N/H]}_{\rm HVC}$ for a given ${\epsilon}_{\rm sf}$.
It should be stressed here that the derived $M_{\rm HVC}$
is for the ISM that forms young stars: it is not for the {\it  entire}
ISM of the LMC.

Fig. 1 also shows that the mass ratio of $M_{\rm HVC}$
to $M_{\rm g}$ is dependent on ${\rm  [N/H]}_{\rm HVC}$
in the sense that a larger amount of HVCs is necessary
to lower the [N/H] of the ISM to the observed level
for larger ${\rm  [N/H]}_{\rm HVC}$. 
The derived large  mass-ratios  $M_{\rm HVC}/M_{\rm g}$ 
(ranging from $\sim 2.3$ to $\sim 5.8$)
mean that a significant degree of dilution of original ISM of the LMC
by HVC infall is indispensable for explaining the observed low
[N/H] in HII regions and young stars in the LMC.
For example, the original gas of the LMC 
with $M_{\rm g}=7.8 \times 10^5 {\rm M}_{\odot}$ 
is converted into new stars for ${\epsilon}=1.0$ 
and ${\rm [N/H]}_{\rm HVC}=-2.0$ after being mixed with
the HVCs with $M_{\rm HVC}=1.8 \times 10^6 {\rm M}_{\odot}$. 
Fig. 2 shows that the required 
$M_{\rm HVC}$ is quite large ($\sim 6.3 \times 10^7 {\rm M}_{\odot}$)
if young stars with ages less than $10^8$yr uniformly have low [N/H]
of $-0.8$. Fig. 2 also shows that the required $M_{\rm HVC}$
depends weakly on ${\rm  [N/H]}_{\rm HVC}$ for a given ${\epsilon}_{\rm sf}$
in models with different $t_{\rm sf}$.

\section{A possible total mass of the HVCs}

Only a small fraction of the Galactic HVCs can interact with the LMC
owing to the small disk size of the LMC.  We here estimate (i)
a typical timescale for the LMC to collide with one HVC ($t_{\rm col}$)
for a given
number density of the HVCs within the distance of the LMC
from the Galactic center 
and (ii) an expected accretion rate of HVCs onto the LMC disk
(${\dot{M}}_{\rm HVC}$).
Since we can estimate the accretion rate required for explaining
the observed [N/H] using the results shown in Figs. 1 and 2
(i.e., $M_{\rm HVC}/t_{\rm sf}$), we can compare the expected
and the required accretion rates and thereby assess the viability
of the present scenario.

The time scale of a LMC-HVC collision event ($t_{\rm col}$)
can be estimated
as follows (e.g., Makino \& Hut 1997);
\begin{equation}
t_{\rm col}=\frac{ 1 } {n_{\rm HVC}\sigma v},
\end{equation}
where $n_{\rm HVC}$, $\sigma$, and $v$
are the mean number density of the HVCs within
the Galaxy,
the geometrical cross section of the LMC,
and a relative velocity between a HVC  and the LMC.
We here estimate $n_{\rm HVC}$ for the central 75 kpc
of the Galaxy (corresponding roughly to the mean
of the pericenter and apocenter distances  of the LMC orbit;
e.g., Bekki \& Chiba 2005)
and assume that  $\sigma= \pi {R_{\rm LMC}}^2$,
where $R_{\rm LMC}$ is the LMC size
and $v$ is velocity dispersion ($=v_{\rm c}/\sqrt(2)$,
where $v_{\rm c}$ is the circular velocity thus 220 km s$^{-1}$)  of the Galaxy
halo. For convenience, we discuss   $t_{\rm col}$  in terms
of the total number of HVCs within 75 kpc from the Galaxy ($N_{\rm HVC}$)
rather than $n_{\rm HVC}$  below.
Previous observations found more than 600 HVCs   
(Wakker \& van Woerden 1991) yet many initial HVCs have been already  destroyed
by tides and ram pressure (e.g., See Wakker 2004 for a review).
Thus we consider that it is reasonable to
investigate  models with $N_{\rm HVC}$ ranging from a few hundreds
to a few thousands.

Fig. 3 shows that  $t_{\rm col}$ is shorter for larger $N_{\rm HVC}$
for a given $R_{\rm LMC}$ and shorter for larger $R_{\rm LMC}$
for a given $N_{\rm HVC}$.
Fig. 3 also shows that $t_{\rm col}$ can be as low as $\sim 10^8$yr if
the total number of the HVC within the LMC's orbit
is as large  as $\sim 1000$ for the size of the LMC disk ($R_{\rm LMC}=5$ kpc).
It is clear from this Fig. 3
that if there are only a few hundreds  HVCs within the LMC orbit,
then the HVCs are highly unlikely to be accreted onto the LMC and consequently
dilute the ISM within a timescale of well less than $10^8$yr (which corresponds
to ages of young stellar populations with low [N/H]  in the LMC).

By assuming a typical mass of the individual  HVCs ($m_{\rm hvc}$) and
using the results shown in Fig. 3,
we can discuss the possible accretion rate of the HVCs onto the LMC disk
for a given set of model parameters.
Fig. 4 shows that ${\dot{M}}_{\rm HVC}$ is much less than 
$\sim 0.1 {\rm M}_{\odot} {\rm yr}^{-1}$ for almost 
all models with different $m_{\rm hvc}$ and $N_{\rm HVC}$.
The minimum value of the required $M_{\rm HVC}$ shown in Fig. 1
is $1.8 \times 10^6 {\rm M}_{\odot}$ for $t_{\rm sf}=10^7$yr
in different models with different ${\epsilon}_{\rm sf}$ and
${\rm [N/H]}_{\rm HVC}$. Therefore, at least 
0.18 ${\rm M}_{\odot}$ yr$^{-1}$ is necessary to dilute the ISM
of the LMC to the observed level for $t_{\rm sf}=10^7$ yr.
It should be stressed that the above 0.18 ${\rm M}_{\odot}$ yr$^{-1}$
is for ${\epsilon}_{\rm sf}=1.0$ (i.e., 100\% star formation efficiency): 
a realistic value of
the required ${\dot{M}}_{\rm HVC}$  is likely to be  significantly
larger than 0.18 ${\rm M}_{\odot}$ yr$^{-1}$.

The results shown in Fig. 4 suggest that only models
with very large typical masses of HVCs 
(i.e., $m_{\rm hvc} = 10^7 {\rm M}_{\odot}$)
and large number of the HVCs ($N_{\rm HVC}>2500$) can show
${\dot{M}}_{\rm HVC}$ as high as the required rate above
(0.18 ${\rm M}_{\odot}$ yr$^{-1}$).
Although the required typical mass is similar to
the observed mass of  Complex C (e.g., Thom et al. 2008),
the required total number within the LMC's orbital radius
already exceeds the total number of the HVCs ($\sim 2000$) observed
by the HI Parkes All Sky Survey (HIPASS; e.g., Putman et al. 2002):
it should be noted that the observed one is for the HVCs
existing in the entire regions around the Galaxy
whereas the required one is only for those within $\sim 75$ kpc.
These results imply that it is unlikely for the  accretion
of the Galactic HVCs onto the LMC disk  to dilute
the ISM.
However ${\dot{M}}_{\rm HVC}$
could  become large enough in a sporadic way
if the LMC can interact with groups of HVCs with locally
large number densities.

\section{Discussion}

\subsection{On model uncertainties}

Although we have adopted a reasonable range of model parameters
and thereby investigated (i) the accretion rate of the HVCs onto the 
LMC and (ii) the possible total mass of the HVCs within the outer
Galactic halo,  there could be some uncertainties in model parameters.
Thus we here discuss how the present results depend on these model
parameters.

\subsubsection{The required accretion rate of the HVCs}

We  assumed that (i) all of young stellar populations with ages
less than $t_{\rm sf}$ have unusually low [N/H] in the LMC and (ii)
the populations are formed exclusively from mixed gas of the original
ISM of the LMC and the accreted gas. If we relax the first assumption,
then $M_{\rm HVC}$ required for explaining the observed low [N/H]
in the LMC can change significantly. In the following discussion,
we define $f_{\rm sf}$ as a fraction of stars having unusually low
[N/H] among all stars formed during  $t_{\rm sf}$ for convenience.
For example, if only 10\% (i.e., $f_{\rm sf}=0.1$) of
the young populations with $t_{\rm sf} \le 10^7$ yr
can show low [N/H] in models with ${\epsilon}_{\rm sf}=0.3$,
then the required $M_{\rm HVC}$ 
can be by a factor of 10 smaller than those derived in 
models (with  ${\epsilon}_{\rm sf}=0.3$) shown in Fig. 1:
the required $M_{\rm HVC}$ can be derived from the equations (1) and (2)
by replacing $t_{\rm sf}R_{\rm sf}$ by $f_{\rm sf}t_{\rm sf}R_{\rm sf}$.

Given the two equations (1) and (2),  the required $M_{\rm HVC}$
is smaller for smaller $f_{\rm sf}$. This means that  a smaller 
number of HVCs need to be accreted by the LMC so that the observed
unusually low [N/H] can be explained by the HVC accretion/collision
scenario. This furthermore  means that a smaller number of HVCs need
to exist in the outer Galactic halo for a given
typical mass of the HVCs  (see discussion in \S 3).
However, young stellar populations 
are observed to  have a low dispersion in [N/H]
(e.g., Russel \& Dopita 1990): it is highly unlikely that
a significant fraction of young stellar populations
have normal [N/H] (i.e., $f_{\rm sf}$ can be close to 1,
as adopted in \S 2.).

If we relax the second assumption (i.e., only some fraction
of the HVC can be mixed into the ISM of the LMC),  then the required
$M_{\rm HVC}$ can also change. In the following discussion,
we define $f_{\rm HVC}$ as a mass fraction of HVCs that can be
mixed into ISM and then converted into new stars for convenience.
Given the equations (1) and (2),  
the required $M_{\rm HVC}$ can be larger for smaller $f_{\rm HVC}$
(the required $M_{\rm HVC}$ is inversely proportional
to $f_{\rm HVC}$). This means that a larger number of the Galactic HVCs
need to exist in the outer Galactic halo for smaller $f_{\rm HVC}$.
We adopted $f_{\rm HVC}=1$ in \S 2 and 3 and showed
that the required number of the HVCs appears to be already too large.
Thus the HVC accretion/collision
scenario becomes less viable if we adopt  smaller $f_{\rm HVC}$:
the main conclusion that the HVC scenario is unlikely (as described later)
does not depend on $f_{\rm HVC}$.

\subsubsection{The possible total mass of the HVCs}

Even if our estimation of the required $M_{\rm HVC}$ is reasonable,
there could be some model uncertainties in estimating the total
mass of the HVCs in the outer Galactic halo and the possible
accretion rate of the HVCs onto the LMC.
Given the equation (3),  $t_{\rm col}$ 
(thus ${\dot{M}}_{\rm HVC}$) can change
if we adopt different $v_{\rm c}$.  For example,
$t_{\rm col}$ is by a factor of 0.88 smaller if we adopt
$v_{\rm c}=250$ km s$^{-1}$ that has been recently suggested
by observations (e.g., Uemura et al. 2000).
This means that ${\dot{M}}_{\rm HVC}$ can be by a factor
of $1.1$ larger in models with $v_{\rm c}=250$ km s$^{-1}$ 
than those in models with  $v_{\rm c}=220$ km s$^{-1}$ (shown
in Fig. 4). This very small  change of ${\dot{M}}_{\rm HVC}$
due to possibly different $v_{\rm c}$ suggests that
the present results on ${\dot{M}}_{\rm HVC}$ can be reasonable.

It should be stressed here that the uniform distribution of the HVCs
within the Galactic halo is assumed in the present estimation.
Therefore, it would be possible for  ${\dot{M}}_{\rm HVC}$
to become large enough in a sporadic way
if the LMC can interact with groups of HVCs with locally
large number densities.
It is however very hard to estimate this effect of sporadic accretion
in a quantitatively way owing to lack of observational results on the 3D
distribution of the Galactic HVCs.
Therefore we can just say that ${\dot{M}}_{\rm HVC}$
depends on the 3D spatial distribution of the HVCs within
the Galactic halo and thus that
the present study can underestimate ${\dot{M}}_{\rm HVC}$
significantly.

\subsubsection{Delayed star formation after gas infall ?}

We have so far assumed that star formation can occur immediately after
the collision/accretion of the Galactic HVCs onto the LMC's gas disk.
Owing to this assumption, a larger amount of HVCs needs to be accreted
within a relatively short time-scale ($\sim 10^7$ yr). However,
if star formation events due to the HVC collision/accretion can be well
(e.g., $\sim 10^8$ yr) after the HVC accretion events
(and if only the very young stars have unusually low [N/H], as assumed
in the present paper),
then the required rate of the HVC accretion can become significantly lower:
the required total mass of the accreting HVCs is the same, but
the HVCs can be accreted within  a longer time scale so that
the net accretion rate can be significantly lower.

Given that $\sim 10^8$ yr corresponds  to  one rotation period
of the LMC for a reasonable set of dynamical parameters of the LMC
(e.g., Bekki \& Chiba 2005),
{\it global mixing}  of the ISM and infalling gas due to kpc-scale dynamical
processes (e.g., dynamical action of the stellar bar) can happen
within $\sim 10^8$ yr.
This means that
the proposed  delayed star formation 
would be promising only if the accreted HVCs
do not {\it globally}  mix with the almost entire  ISM with normal [N/H] for
a such long timescale of $\sim 10^8$ yr,
because such global chemical mixing will result in larger [N/H] of stars formed
well after the HVC accretion events owing to the much larger total  mass
of the ISM.

\subsection{Infall from the SMC rather than from the HVCs ?}

We adopted an assumption that only very young stellar populations
formed from mixed gas of HVCs and ISM in the LMC
have unusually low [N/H] in the LMC:
the accretion events of HVCs onto the LMC need to happen
only  recently.
We have shown that the large number ($> 2500$) of massive
HVCs ($\sim 10^7 {\rm M}_{\odot}$) are required to exist 
within the LMC's orbital radius with respect to the Galactic center:
the required total mass of HVCs ($M_{\rm HVC, G}$) in the Galactic halo
is about $\sim 2.5 \times 10^{10} {\rm M}_{\odot}$ for a reasonable set of 
model parameters.
Although the previous numerical simulations tried to predict the total
mass of the Galactic HVCs,
the predicted mass ranges widely from $\sim 10^8 {\rm M}_{\odot}$ 
(Peek et al. 2008) to $\sim 2 \times  10^{10} {\rm M}_{\odot}$ 
(Maller \& Bullock 2004).
The required $M_{\rm HVC, G}$ 
to explain the observed low [N/H] in the present scenario
appears to exceed the predicted $M_{\rm HVC, G}$.

Given that the  typical mass and 3D distribution of the Galactic HVCs
remains observationally unclear (e.g. Wakker 2004),
the above inconsistency between the required total mass of HVCs
and the theoretically predicted one does not rule out the present
scenario.
It is, however, reasonable that gas from other sources (e.g., galaxies
in the Local Group) can also play a role in diluting the ISM of the LMC.
Recently, Bekki \& Chiba (2007) have shown that the ISM  stripped from 
the SMC during the LMC-SMC-Galaxy interaction for the past 2 Gyr can
collide with the LMC's disk around 0.2 Gyr ago. 
We thus suggest the following
``SMC-transfer'' scenario 
(or ``Magellanic squall''; Bekki \& Chiba 2007).
During the last 0.2 Gyr, 
the LMC and the SMC have interacted each other like a binary
through their strong gravitational fields. 
As a result of this tidal interaction,  gas 
with low [N/H] in the SMC can be  transferred 
efficiently to the LMC sporadically, 
which induces 
star formation and thus creation of HII regions
with low [N/H] in the LMC.
Thus the observed low [N/H] 
of young populations in the LMC  is a  result of a close
tidal  interaction between MCs 
in the last 0.2 Gyr or so.

We here suggest that this SMC-transfer scenario 
has the following three advantages in 
explaining the observed low [N/H].
Firstly, the relative velocities between the infalling gas from the SMC
and the LMC's gas disk can be as small as $\sim 60$ km s$^{-1}$,
because the relative velocity between the LMC and the SMC
is $\sim$ 60 km s$^{-1}$ for the last 200 Myrs (e.g., orbital models
of the LMC and the SMC
shown in Bekki \& Chiba 2005). The relative velocity is significantly
smaller than the circular velocities of the LMC ($\sim 80-120$ km s$^{-1}$
for a reasonable mass model of the LMC; Bekki \& Chiba 2005)
so that the infalling gas is highly likely to be trapped by the gravitational
potential of the LMC.
On the other hand, the relative velocities of the HVCs and the LMC
can be as large as the velocity dispersion of the Galactic halo
($\sim 160$ km s$^{-1}$) so that the infalling HVCs are less likely
to be trapped by the LMC in comparison with the infalling SMC gas.

Secondly Bekki \& Chiba (2007)
have shown that
about 18\% of the gas within  the SMC
can pass through 
the LMC about 0.2 Gyr ago.
If the SMC's initial gas mass before
gas stripping  is $\sim 10^9 {\rm M}_{\odot}$,
then a significant amount of gas (as much as  $\sim 10^8 {\rm M}_{\odot}$)
can be accreted onto the LMC.
This is much larger than the total amount of HVCs that can be accreted onto
the LMC for the last $\sim 10^7$  yr, as shown in the previous sections.
Furthermore, the accretion event of a large amount of gas from the SMC
can occur when the SMC approaches the LMC very closely so that
the accreted gas can {\it simultaneously} trigger star formation in the LMC: most
HII region can show systematically low [N/H].
Possibly sporadic accretion of HVCs would be unlikely to cause such
synchronized star formation in the LMC.

Thirdly, Bekki \& Chiba (2007) have already shown that
the gas-transfer between the LMC and the SMC is possible for the last 200 Myrs
using the results of numerical simulations. 
However, no one has demonstrated that the massive HVCs like Complex C with
physical sizes of 10 kpc $\times$ 10 kpc (e.g., Wakker et al. 1999) can be
really accreted onto the LMC owing to hydrodynamical interaction
between the LMC's gas disk and the HVCs in spite of the large relative velocities
($\sim 160$ km s$^{-1}$)  between them. 
If only some minor fractions (e.g., 10\%) of the HVC masses can be accreted
onto the LMC during HVC-LMC collisions, then the required number of the Galactic
HVCs for explaining the observed [N/H] in the LMC
can be unrealistically large.

Thus, if the ISM of the SMC has rather low [N/H] 
and if  the stripped gas can be mixed into the ISM of the LMC
and then converted into new stars,
the newly formed stars  can show low [N/H]. 
Indeed the HII regions and 
young stellar populations of the SMC are observed to have
[N/H] by a factor of $\sim 18$ lower than
the solar value  (e.g., Pilyugin et al. 2003;
 Rolleston et al. 2003; Hill 2004),
which implies that the ISM of the SMC
could possibly have low [N/H] (though the low [N/H]
could be only for the young stellar populations, not for
the entire ISM).
The required gas corresponding to this [N/H]
is $\sim 10^6 -10^7 {\rm M}_{\odot}$
(see Fig.1). Thus, the predicted amount of gas
transferred from the SMC to the LMC of $\sim 10^8 {\rm M}_{\odot}$
(Bekki \& Chiba 2007)
is sufficient to dilute the N abundance in the LMC as observed.

However, the SMC-transfer scenario has some disadvantages in
explaining clearly the observed low [N/H] both in the LMC 
and the SMC. For example,
if the origin of the unusually low [N/H] in the LMC is due to
the gas transfer between the Clouds, 
then the next question
is as to why the SMC has ISM with such low [N/H]: this point
is yet to be answered by the SMC-transfer scenario.
Previous chemical evolution models did not clearly show that 
the present-day dwarf galaxies
like the SMC can have very low [N/H]
(e.g., Henry et al. 2000; Moll\'a et al. 2006):
the Appendix also implies  that 
canonical chemical evolution models can hardly show
very low [N/H] in the present stellar populations
for Magellanic-type dwarf galaxies. 
We need to discuss  why the ISM of the SMC can have
low [N/H] in our future paper (Bekki \& Tsujimoto 2010).

Also, the gas infall of such low-metallicity gas 
(with a possibly large  mass of $\sim 10^8 {\rm M}_{\odot}$) from the SMC
would lower [Fe/H] of the LMC significantly:
although
the presence of metal-poor young clusters 
(e.g., NGC 1984 with an age of $\sim 4$ Myr
and [Fe/H]$\sim -0.9$; Santos \& Piatti 2004)
would suggest a possible evidence of dilution of ISM
by low-metallicity gas,
there are no observational results which suggest that
the young stellar populations as a whole show such low [Fe/H]
in the LMC (It should be noted here that
this problem may be true for the HVC scenario).
Furthermore, the gas accretion from the SMC to the LMC can occur 
most efficiently about 0.2 Gyr ago (Bekki \& Chiba 2007).
This  means that the SMC-infall scenario needs to explain
how the gas infall can still continue to occur until quite recently
(until only 10 Myrs ago) so that 
the very young populations of the LMC
can be formed with low [N/H] from the accreted gas.

The HVC infall scenario suggests that if the HVCs can be accreted onto
the LMC, then they can be accreted also onto the SMC owing to
the similar locations and velocities between the LMC and the SMC
with respect to the Galactic center. Therefore, it can naturally
explain why both the LMC and the SMC show low [N/H] in their young
stellar populations. As pointed out above, both the HVC infall scenario
and the SMC-transfer one have advantages and disadvantages in explaining
the observed chemical properties of the LMC and the SMC.
Thus  it would be reasonable for us to say that both scenario
are possible at present.

\subsection{A possible observational evidence for external gas infall
onto the LMC}

If chemical evolution of the LMC is influenced by accretion of gas from
outside the LMC (e.g., from the Galactic halo or other gas-rich galaxies),
then the chemical evolution strongly depends on
the orbital evolution of the LMC.
The LMC may have obtained steadily gas from the accretion events
of the Galactic HVCs
for at least $3-4$ Gyr for the ``classical bound orbit'' 
adopted in previous dynamical models for the evolution of the LMC
(e.g., Bekki \& Chiba 2005).
For this classical orbital model, the LMC can show
lower [N/H] not only in young stellar populations but also
in intermediate-age ones,
if the dilution by the continuous gas infall
has been  overwhelming over chemical
enrichment by the  AGB stars continuously formed  in the LMC
for the last 3-4 Gyr.
If the LMC has just recently arrived in the Galaxy,
as the latest proper motion
studies by {\it HST} (Kallivayalil et al. 2006) has suggested, then
the LMC may have started the accretion of the HVCs quite recently
(well less than $\sim 1$ Gyr): the low [N/H] can be seen only in
young stellar populations.

We have so far considered that all of young stellar populations
in the LMC have
unusually low [N/H] and were  formed from {\it mixed gas} of the LMC's ISM
and the accreted gas from outside.
However,
some local regions where accretion of gas with very low [N/H] 
does not occur may well form  young stellar populations with
normal [N/H]. 
In this case, the young populations in the LMC may well
show a large dispersion in [N/H] in the present scenario. 
The previous observations on the chemical abundances for HII regions
of the LMC indeed shows a dispersion in [N/H] 
(e.g., Russell \& Dopita 1990),
though the dispersion appears to be smaller (see Fig. 5).
If future observations confirm that the dispersion in [N/H] is really small
in the LMC
(i.e., [N/H] is uniformly low for the entire young  populations),
then it means that young  populations in the LMC were formed exclusively
from mixed gas of the original ISM of the LMC  and the accreted gas
for some physical mechanisms.

However, it should be stressed that if the {\it entire ISM}
(i.e., not only cold HI and molecular gas but also HII regions)
of the present LMC
has unusually low [N/H] for some physical reasons,  no accretion
event of gas outside the LMC is required for explaining 
the origin of the observed
young stellar populations with unusually low [N/H]: both old and young
stellar populations have unusually low [N/H].
If this is the case, then the next question is
why and how the [N/H] of the ISM in the LMC has continued
to be very low until  the mean metallicity of the LMC has become
as high as [Fe/H]$\sim -0.3$. 
No previous chemical evolution models appear to have shown
very low [N/H] in {\it the present} galaxies with lower
mean metallicities like the LMC (e.g., Moll\'a et al. 2006). 
This implies that it is unlikely that the entire ISM  of the LMC
has very low [N/H]: it is natural to consider that
external gaseous accretion has recently changed [N/H] in the ISM
with originally normal [N/H] (see the Appendix on this issue).

Then, is there any possible observational 
evidence for recent accretion events of gas
in the LMC? In order to answer this question, we have investigated
the dependences of $\log$(N/O) on 12+$\log$(O/H) for the HII regions
of the LMC using already existing data sets 
(Peimbert \& Torres-Peimbert 1974;
Dufour 1975;
Pagel et al. 1978;
Russell \& Dopita 1990;
Simpson et al. 1995).
Fig. 5 shows that (i) there is no clear correlation between
$\log$(N/O) and 12+$\log$(O/H)
and (ii) HII regions with larger 12+$\log$(O/H) do not show
larger $\log$(N/O).
These are  inconsistent with previous chemical evolution models
(e.g., Henry et al. 2000) 
demonstrating that  stars and gas with  larger 12+$\log$(O/H)
have larger $\log$(N/O): It should be noted here that infall of gas
with low [N/H] are not included in the models.
We thus suggest that the observed 
 lack of HII regions with larger 12+$\log$(O/H)
and larger $\log$(N/O) are due to accretion of gas with low [N/H]:
original  HII regions with larger 
$\log$(N/O) and larger 12+$\log$(O/H) can disappear owing to
the dilution of [N/H] caused by the gaseous accretion.

\subsection{Observational implications: effects on other
elements}

Recent gas infall should also influence abundances of young population
in the LMC for the elements other than N. Here we discuss this issue by
comparing the LMC abundance with those of HVCs or the SMC.
Supergiants in both the LMC and the SMC basically exhibit the solar abundance
ratios for $\alpha$-elements and iron-peak elements (Hill 2004). In addition, both
MCs exhibit a similar overabundance for neutron-capture elements (Hill 2004).
In this way, two galaxies have a very similar present-day elemental abundance pattern.
It implies that  gas infall from the SMC would dilute the N abundance with little imprint in
other elemental ratios.

On the other hand, the information on abundances of HVCs for the
 elements other than N and O is very restricted. 
Complex C seems to exhibit an essentially solar pattern for
 $\alpha$-elements  (O, S, Si) in comparison with
 Fe but with a possible slightly-enhanced 
[$\alpha$-elements/Fe] (Collins et al. 2007).
 If the future observation reveals the clear 
SN II-like [$\alpha$-elements/Fe] for HVCs, we will 
be able to  conclude that the dilution by 
them will change the LMC abundance into the one at odds 
with the observed LMC abundance. 
However, at the moment, it can be said that
large impact on abundances by  infall of HVCs is 
likely to be seen only in the deficiency of the N abundance.

\subsection{Other possible scenarios}

The proposed accretion scenario in the present paper
might well be   one
of possible scenarios.  Whatever alternative scenario is proposed,
it would need to discuss the origin of the unusually low [N/H]
in the context of an unique LMC's  environment (e.g., interaction
with the Galaxy and the SMC).
We here discuss three alternative scenarios that could possibly
explain the observed unusually low [N/H].
The first  is that 
the stellar winds of AGB stars (i.e., rich sources of nitrogen)
can be efficiently and continuously stripped from
the LMC owing to hydrodynamical interaction between the
LMC and the Galactic warm halo gas so that the present [N/H]
can be rather low.
This scenario with selective loss of AGB ejecta
would have a problem of explaining other observational results
on chemical abundances of stars (e.g., the
observed abundances  of s-process elements).

However, it is possible that
stellar ejecta only from massive AGB stars (i.e., rich source
of nitrogen)
are removed owing to the stronger winds whereas  those from low-mass ones
(i.e., rich sources of s-process elements)  are not.
If this is the case, the above problem associated with
the s-process elements would not be so serious  for the
selective loss scenario.
We consider that this scenario is highly unlikely,
because
it  needs to explain why
gas from supernovae (which are more energetic)
can be kept within 
the LMC: recycling of the supernovae ejecta  is inevitable  to reproduce
the chemical feature of the LMC.

The second is that the initial mass function (IMF) is different in
the LMC in the sense that a much smaller number of  stars
(i.e., massive  AGB stars)
mainly contributing to  the production of nitrogen
can be formed in the history of the LMC. 
This idea confronts the same problem as the first one. 
The LMC requires the IMF that keeps the number of both low-mass 
AGB stars and massive ($>10 {\rm M}_{\odot}$) stars while reduces the 
number of stars residing in the middle of them. It is hard to formalize 
such an IMF. 
The third one is that nitrogen production in AGB stars is significantly 
suppressed in the LMC at least for the last few Gyr.
 Indeed, the observed level of 
N enrichment in the solar neighborhood
can not be realized without introducing the 
metallicity-dependent N yield, 
which  is also predicted by theoretical studies 
on nucleosynthesis in AGB stars (e.g., Ventura et al. 2002). Therefore, 
if the N yield did not depend on the metallicity, 
the present-day N abundance 
would become significantly 
smaller as observed in the LMC. Evidently, 
there is no physical reason for making such difference between the LMC and 
the Galaxy.

Thus, the three alternative scenarios are much less convincing 
in comparison with the present one in which the dilution of
ISM by gas infall is responsible for the observed low [N/H] in
young stellar populations of the LMC, 
though we did not quantitatively
investigate  the evolution of nitrogen abundances in stars and ISM
of the LMC using chemical evolution models with the
proposed unusual IMFs.
The required gas with low [N/H] is less likely to come from the Galactic HVC,
though the HVC infall scenario can not be ruled out currently. 
The gas accretion  from the SMC is a possible scenario, which however
has both advantages and disadvantages in explaining the observations.
We suggest that
the observed unusually low [N/H] in the LMC seems to  tell us 
something about the unique chemical evolution history 
coupled with the past dynamical evolution.

\section{Conclusion}

Given that the observed unusually low [N/H]
in young populations of the LMC can not be simply reproduced
by canonical chemical evolution models for the LMC,
we have investigated
a new scenario in which
the observed  low [N/H] 
can be closely associated with recent collision and subsequent
accretion  of
HVCs that surround the Galaxy and
have   low nitrogen abundances.
We have shown  that even if the observed low [N/H] is limited to
very young stars with ages less than $\sim 10^7$ yr,
the collision/accretion rate of the HVCs onto the LMC
needs to be $\sim 0.2 {\rm M}_{\odot}$ yr$^{-1}$
(corresponding to the total HVC mass of $10^6-10^7 {\rm M}_{\odot}$)
to dilute
the original interstellar medium (ISM) before star formation.

We have demonstrated  that
if the typical mass
of HVCs accreted onto the LMC is $\sim 10^7 {\rm M}_{\odot}$,
the Galaxy needs to have  $\sim 2500$ massive  HVCs within the LMC's
orbital radius with respect to the Galactic center in order
to explain the required accretion rate of the HVCs.
Although we have adopted a number of model assumptions in 
deriving the required number of the HVCs,
the required  number of massive HVCs
suggests that
the HVC infall scenario is possible yet unlikely to efficiently dilute the ISM
of the LMC and consequently lower the  [N/H].
We thus have discussed  the alternative SMC-transfer scenario in which 
the transfer of gas with low [N/H] from the SMC
to the LMC can  explain
the observed low [N/H].

The SMC-transfer scenario has some
advantages in explaining the observations
(e.g., a higher probability of an enough amount
of  gas to be accreted onto the LMC) over the HVC infall one.
However it also appears to have  some
problems in explaining self-consistently the observed chemical properties
of the LMC. For example, it is not so clear in the SMC-transfer scenario why
the gas accretion can continue to occur until quite recently.
We thus conclude that although the observed low [N/H] of young populations
in the LMC has an external origin (i.e., gas infall from outside the LMC),
the host objects which the external gas
originates from  are  yet to be determined.

\acknowledgments
We are  grateful to the anonymous referee for valuable comments,
which contribute to improve the present paper.
K.B. acknowledges the financial support of the Australian Research
Council throughout the course of this work.
TT is assisted in part by Grant-in-Aid for Scientific Research
(21540246) of
 the Japanese Ministry of Education, Culture, Sports, Science and
Technology.

\section{The Appendix}

We consider that it is useful to show how the nitrogen
abundance in the LMC  evolves
with time in a reasonable one-zone chemical evolution model of the LMC,
because no theoretical works have yet clearly demonstrated that
canonical chemical evolution models can not reproduce the observed 
low [N/H] in the LMC.
We here discuss whether the observed [N/O] in the LMC can be reproduced
by a canonical one-zone chemical evolution model which is consistent
with the observed age-metallicity relation of stellar population in the LMC.
Here we assume the standard Salpeter IMF. Details of model description
is
presented in Tsujimoto \& Bekki (2010), including the chemical yields
from
stars with different masses. 
We will discuss time evolution of [N/H] in different galaxies with different
physical properties using these models in our future papers.

Fig. 6 shows both the time evolution of [Fe/H] and [N/O] for the last $\sim 13.5$ Gyr
and the observed values
so that we can clearly see how serious the observed low nitrogen abundance is 
in the standard chemical evolution models with no external infall of gas.
As shown in Fig. 6,  [N/O] can be low in the early stage of the LMC evolution
when ${\log}$(N/O)+12 was lower than 8 (when the LMC was much younger).
However it is clear that $\log$(N/O) ($\sim -1.1$) in the present LMC 
(at $\log$(O/H)+12$\sim 8.4$) for the standard chemical
evolution model is much larger than the observed one ($\sim -1.5$).
This result demonstrates that some additional factors such as
unique IMFs and external gaseous infall from outside of the LMC
need to be considered in order to reproduce the observed low [N/O]
in the LMC. 
We suggest that
HVCs and gaseous components  of the SMC can be the promising
candidates for the external infall of gas.

\begin{figure}
\epsscale{0.8}
\plotone{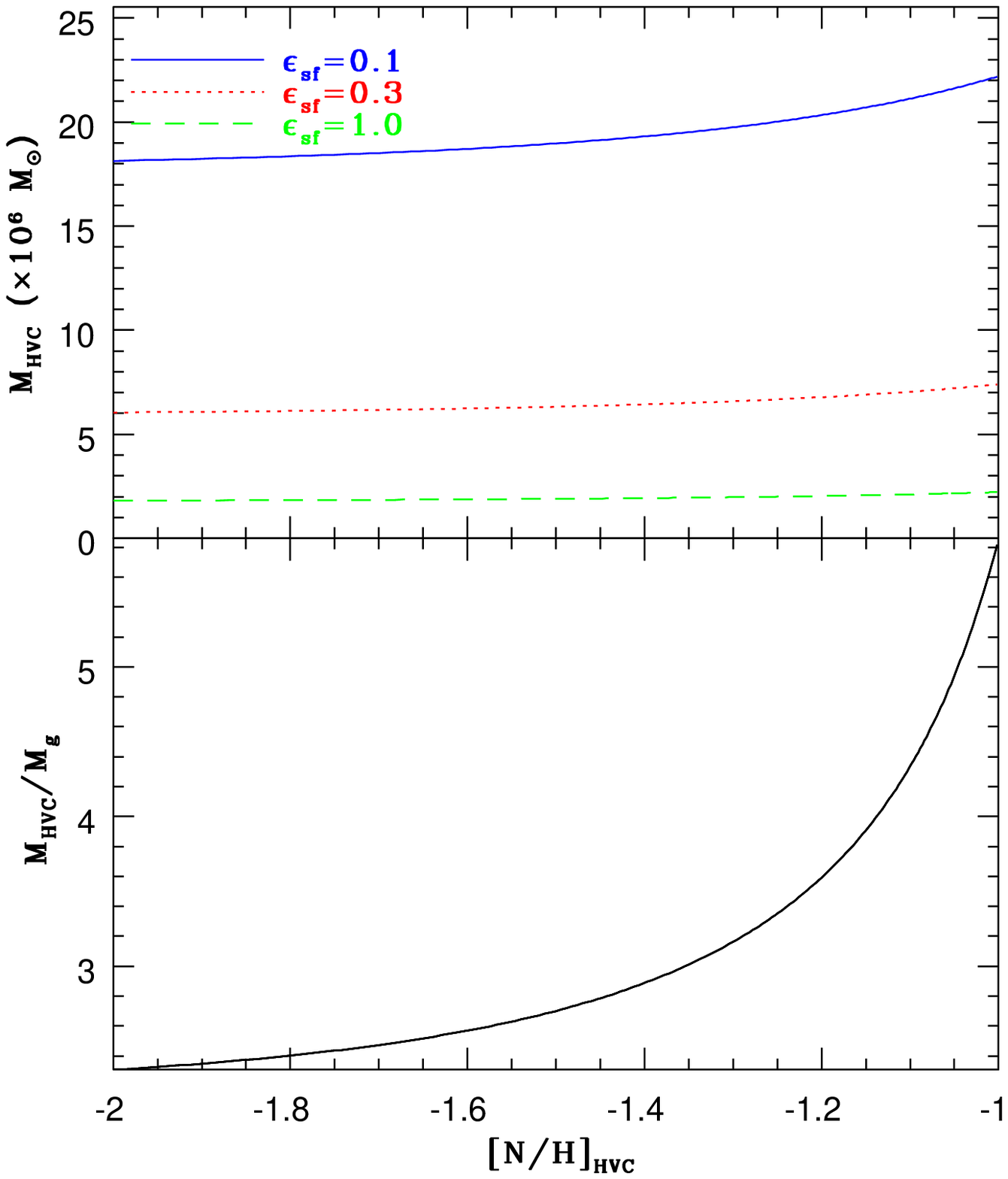}
\figcaption{
The dependences of
$M_{\rm HVC}$ (upper)
and the mass ratio of 
$M_{\rm HVC}$ to $M_{\rm g}$ (lower)
on the nitrogen abundances of the HVCs 
(${\rm [N/H]}_{\rm HVC}$) for three different models with
${\epsilon}_{\rm sf}=0.1$ (blue, solid)
${\epsilon}_{\rm sf}=0.3$ (red, dotted) and
${\epsilon}_{\rm sf}=1.0$ (green, dashed)
for a fixed $t_{\rm sf}$ (=10$^7$ yr).
Here $M_{\rm HVC}$ is the total mass of the Galactic HVCs
required to explain the observations and 
$M_{\rm g}$ is the total mass of the LMC ISM that can mix with
the HVCs to form new stars.
The mass ratio $M_{\rm HVC}/M_{\rm g}$ does not depend
on ${\epsilon}_{\rm sf}$ so that only a line (black, solid)
is shown in the lower panel.
It is clear that a larger amount of HVCs is necessary to
dilute the ISM of the LMC to the observed level
for smaller ${\epsilon}_{\rm sf}$.
\label{fig-1}}
\end{figure}

\begin{figure}
\epsscale{0.8}
\plotone{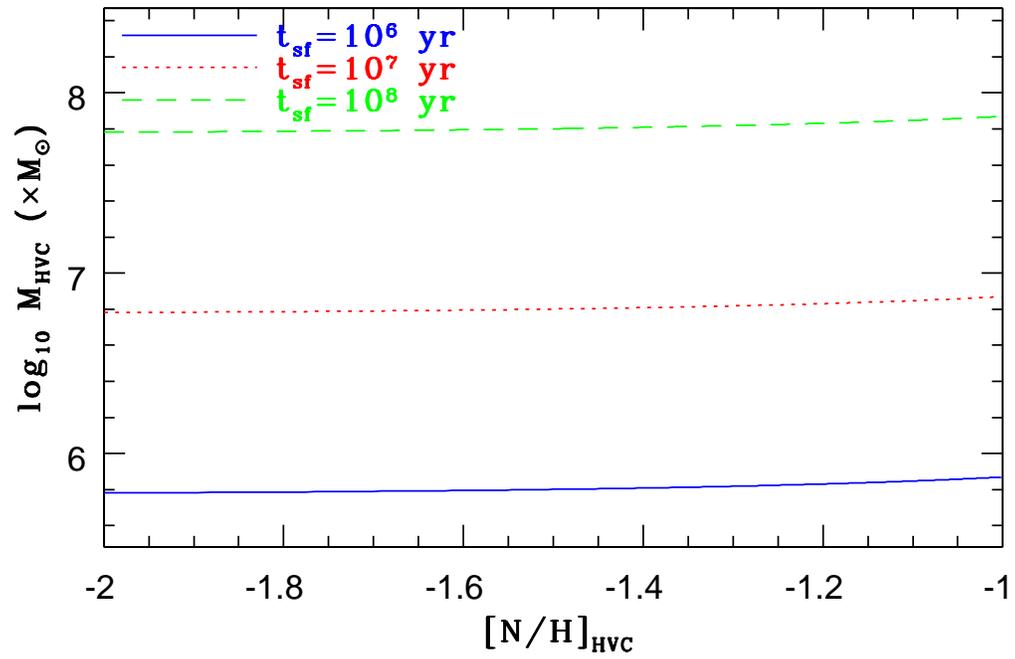}
\figcaption{
The dependences of $M_{\rm HVC}$ 
(in logarithmic scale) on 
${\rm [N/H]}_{\rm HVC}$ for models with 
$t_{\rm sf}=10^6$ yr (blue, solid),
$10^7$ yr (red, dotted),
and $10^8$ yr (green dashed)
for a fixed ${\epsilon}_{\rm sf}$ (=0.3).
\label{fig-2}}
\end{figure}

\begin{figure}
\epsscale{0.8}
\plotone{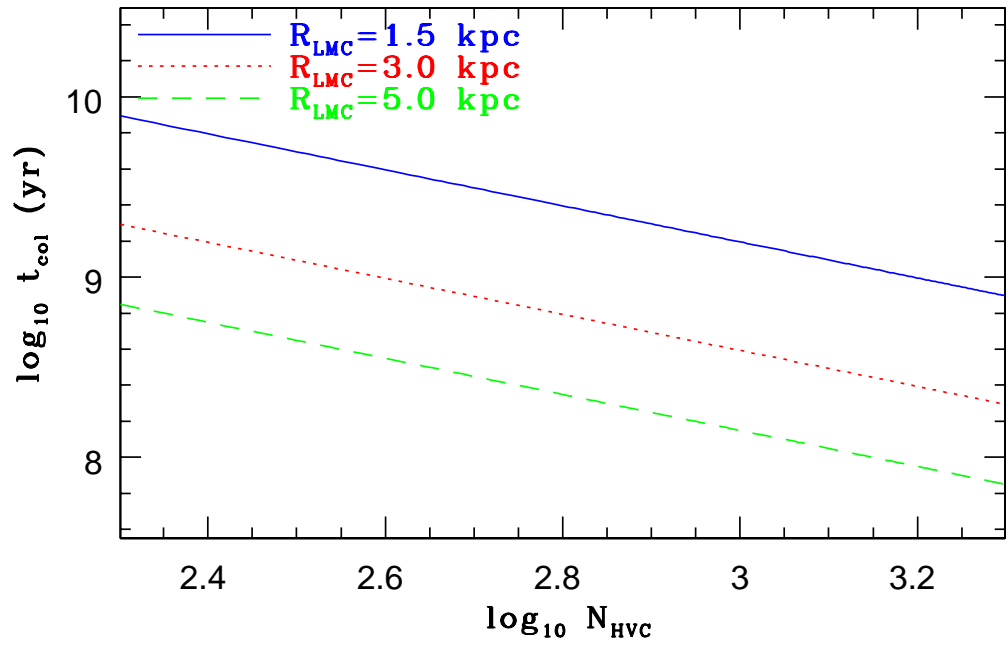}
\figcaption{
The dependences of $t_{\rm col}$ (the timescale of the LMC to collide
a HVC in the Galactic halo) on $N_{\rm HVC}$ (the total number of
the HVCs) for models with $R_{\rm LMC}=1.5$ kpc (blue, solid),
3.0 kpc (red, dashed), and 5.0 kpc (green, dashed).
\label{fig-3}}
\end{figure}

\begin{figure}
\epsscale{0.8}
\plotone{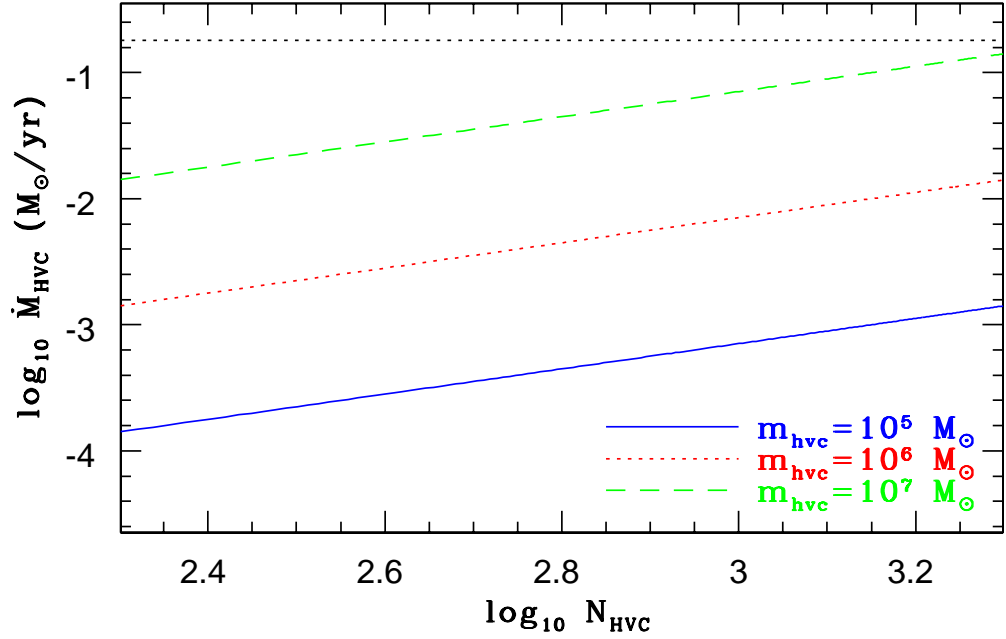}
\figcaption{
The dependences of 
${\dot{M}}_{\rm HVC}$ (the accretion rate of the HVCs onto the LMC)
on $N_{\rm HVC}$ (the total number of the HVCs) for models
with $m_{\rm hvc}$ (the typical mass of individual HVCs) of
$10^5 {\rm M}_{\odot}$ (blue, solid),
$10^6 {\rm M}_{\odot}$ (red, dotted),
and $10^7 {\rm M}_{\odot}$ (green dashed).
A horizontal black dotted line indicates the minimum value of $M_{\rm HVC}$
for models with $t_{\rm sf}=10^7$ yr shown in Fig. 1. 
Note that  ${\dot{M}}_{\rm HVC}$ can not be as high
as the required  $M_{\rm HVC}$ in all three models 
for the adopted  range of $N_{\rm HVC}$.
\label{fig-4}}
\end{figure}

\begin{figure}
\epsscale{0.8}
\plotone{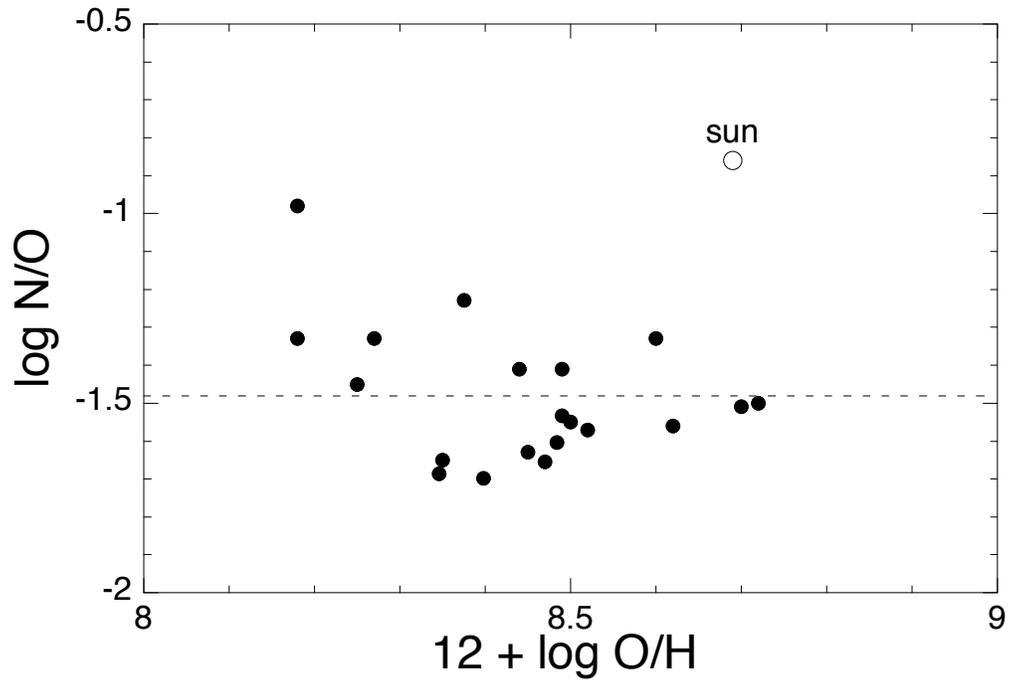}
\figcaption{
The observed dependence of $\log$N/O on 12+$\log$O/H for the HII
regions of the LMC. For comparison, the result for the Sun 
is shown. A dotted  line indicates the mean of the observed HII regions
in the LMC.
\label{fig-5}}
\end{figure}

\begin{figure}
\epsscale{0.8}
\plotone{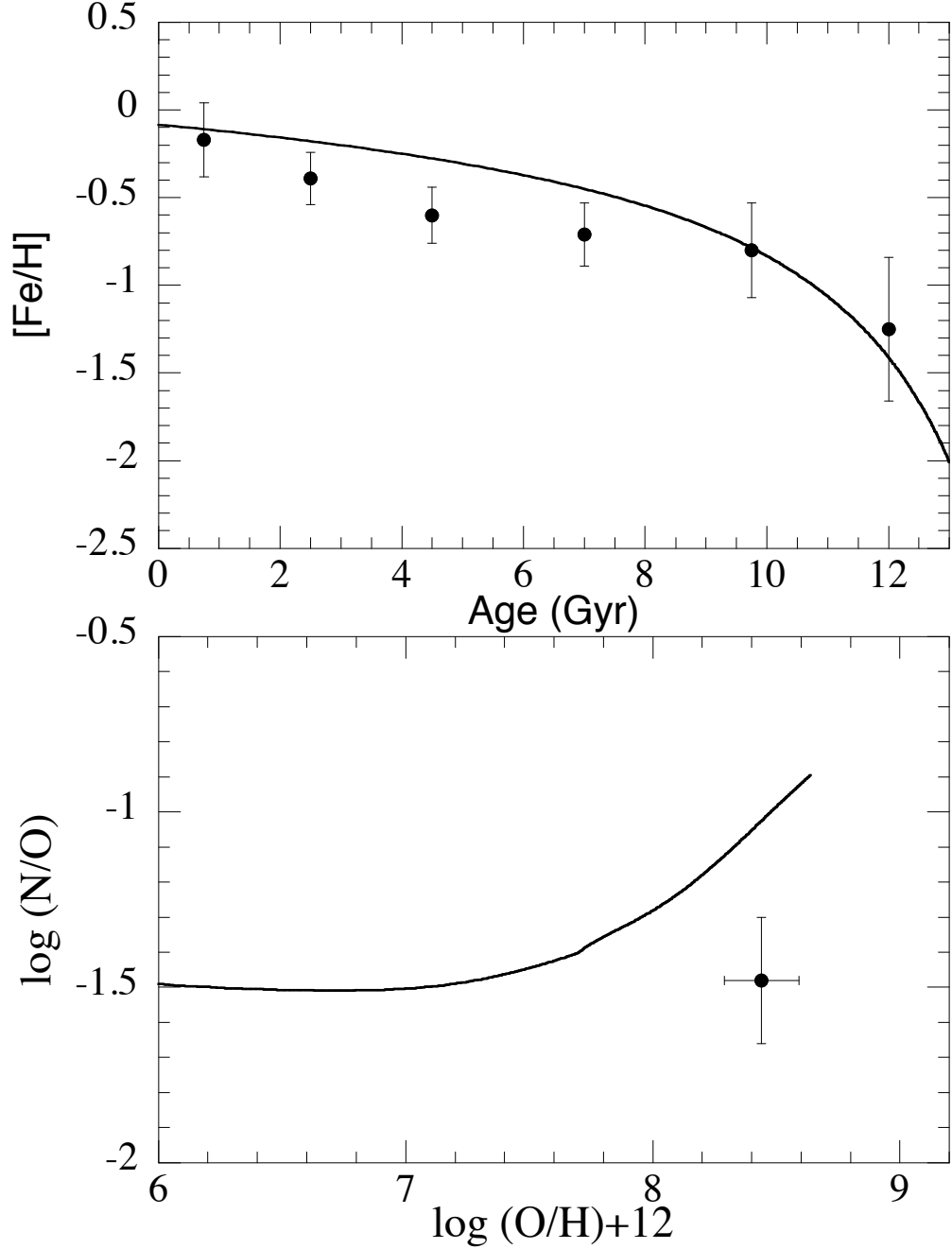}
\figcaption{
The upper panels show the observed age-metallicity relationship of stars in the LMC 
(filled circles with error bars) and
the simulated one based on the standard one-zone chemical evolution
model (solid line).  The lower panel shows the time evolution
of $\log$(N/O) as a function of $\log$(O/H)+12 for the adopted one-zone
chemical evolution model with a reasonable IMF.
The lower $\log$(O/H)+12 means younger LMC in this model.
For comparison, the observed value is shown as a filled circle with error bars.
The details of the observations and the simulated models are given
in Tsujimoto \& Bekki (2009, 2010). Clearly
the modeled present LMC (at $\log$(O/H)+12$\sim 8.4$) shows much larger $\log$(N/O)
in comparison with the observation.
\label{fig-6}}
\end{figure}

\end{document}